\documentclass[preprint,showpacs,preprintnumbers,amsmath,amssymb]{revtex4}

\usepackage{graphicx}
\usepackage{dcolumn}
\usepackage{bm}

\begin{document}

\title{ Optical clocks based on ultra-narrow three-photon resonances in
alkaline earth atoms }

\author{Tao \surname{Hong}, Claire \surname{Cramer}, Warren
\surname{Nagourney}, E. N. \surname{Fortson}}

\address{Department of Physics, University of Washington, Seattle,
Washington 98195.}


\begin{abstract}

A sharp resonance line that appears in three-photon transitions between
the $^{1}S_{0}$ and $^{3}P_{0}$ states of alkaline earth and Yb atoms
is proposed as an optical frequency standard. This proposal permits the
use of the even isotopes, in which the clock transition is narrower than
in proposed clocks using the odd isotopes and the energy interval is not affected by external
magnetic fields or the polarization of trapping light.  The method has
the unique feature that the width and rate of the clock transition can be
continuously adjusted from the $MHz$ level to sub-$mHz$ without loss of
signal amplitude by varying the intensities of the three optical beams.
Doppler and recoil effects can be eliminated by proper alignment of the
three optical beams or by point confinement in a lattice trap. 
The three beams can be mixed to produce the optical frequency corresponding to the $^{3}P_{0}$ - $^{1}S_{0}$ clock interval.

\end{abstract}


\maketitle

Atomic clocks using optical transitions with high intrinsic line-Q offer
unprecedented opportunities for improved metrology standards \cite{Metrology} and tests
of fundamental physics \cite{Significance}.  Recently, much attention
has focused on using the forbidden $^{1}S_{0}$ $\rightarrow$
$^{3}P_{0}$ transitions in alkaline earth and Yb atoms, which become
weakly allowed in the odd isotopes through the hyperfine interaction of the
nuclear spin \cite{SrYb}.  The atoms can be confined in an optical
lattice trap to eliminate the first-order Doppler and recoil effects,
and the trap wavelength can be set to the `magic' value at which the
ground and excited states undergo the same light shift, leaving the
clock transition unshifted \cite{SrYb}.

Here we discuss the alternative of basing a clock on the sharp lines
that appear in three-photon transitions between the $^{1}S_{0}$ and
$^{3}P_{0}$ states.  
This method permits the use of the even isotopes,
in which the clock transition is narrower than in the odd isotopes and
the energy interval is not affected by external magnetic fields or
the polarization of trapping light at the magic wavelength.
The three-photon
scheme offers other interesting options.  By varying the intensities of
the three optical beams the rate, 
and hence the width, of the clock transition can
be continuously adjusted from the $MHz$ level to sub-$mHz$ without loss
of signal amplitude.  Furthermore, proper alignment of the three beams
can eliminate Doppler and recoil effects without the point confinement
of a lattice trap. Finally, the three beams can be mixed to produce the
clock frequency corresponding to the $^{3}P_{0}$ - $^{1}S_{0}$ interval.

\begin{figure}[h]
\includegraphics[width=8cm, height=10cm, angle=270]{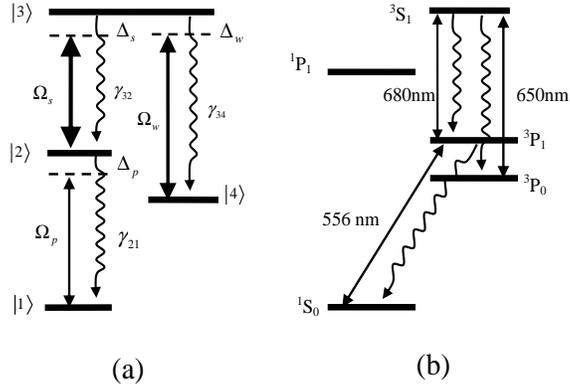}
\caption{ (a) Energy level structure and optical couplings of the
four-level atomic system for making an atomic optical frequency
standard; (b) Specific case of Yb as an example for the scheme in
(a).}
\label{fig1}
\end{figure}

The proposed frequency standard consists of four atomic energy levels
interacting with three light fields: a strong coupling field, a weak
coupling field, and a probe field, as shown in Fig.~\ref{fig1} (a). Our
general results will be applicable to any of the alkaline earth atoms
and Yb, but as an example the four levels for the specific case of
Yb are shown in Fig.~\ref{fig1} (b). Relevant transitions and wavelengths for Ca, Sr and Yb are shown in Table \ref{tab:table1}.
  The idea is based on the concept of
electromagnetically induced transparency and absorption (EITA)
\cite{EITA}.  The three light fields connect the state $|1\rangle$ (the
$^{1}S_{0}$ ground state) to the state $|4\rangle$ (the metastable
$^{3}P_{0}$ state) via two short-lived intermediate states,
$|2\rangle$ and $|3\rangle$ ($^{3}P_{1}$ and $^{3}S_{1}$), but the
width and position of the three-photon EITA features are determined by
the narrow initial and final states and not the relatively broad
intermediate states. In the case of the even isotopes
trapped in an optical dipole trap \cite{SrYb}, the ground and the
metastable state both have zero total angular momentum, so their energy
difference is unaffected by external magnetic fields or trapping light
polarization. Here for simplicity we assume only one sublevel in each intermediate
state ($^{3}P_{1}$ and
$^{3}S_{1}$) participates, and ignore any effect of polarization of the
light fields.  We also ignore other decays of $^{3}S_{1}$, including those to the metastable $^{3}P_{2}$ state, from which atoms can be removed by a separate laser.

Under the rotating wave approximation, the four-level atomic system
coupled to the three light fields can be described by the following
density matrix equation,

\begin{equation}
\frac{d\rho(t)}{dt}=-\frac{i}{\hbar}\left[H,\rho(t)\right]+L\left[\rho(t)\right],
\label{eq1}
\end{equation}
where $\rho(t)$ is the atomic density matrix. The summation of the
diagonal elements satisfies the probability normalization, i.e.,
$\rho_{11}+\rho_{22}+\rho_{33}+\rho_{44}=1$.

The matrix for the system Hamiltonian in the interaction picture is
defined by
\begin{equation}
  H=\hbar\left[\begin{array}{cccc}
      0 & -\Omega_{p}/2 & 0 & 0 \\
      -\Omega_{p}^{*}/2 & -\Delta_{p} & -\Omega_{s}/2 & 0 \\
      0 & -\Omega_{s}^{*}/2 & -\Delta_{s}-\Delta_{p}& -\Omega_{w}/2 \\
      0 & 0 & -\Omega_{w}^{*}/2 & \Delta_{w}-\Delta_{s}-\Delta_{p}
      \end{array}\right],
\label{eq2}
\end{equation}
where $\Omega_p$, $\Omega_{s}$ and $\Omega_{w}$ (see Fig. \ref{fig1}) are the complex Rabi
frequencies associated with the couplings of the probe field, the strong
coupling field and the weak coupling field to atomic transitions
$|1\rangle \rightarrow|2\rangle$ , $|2\rangle \rightarrow|3\rangle$  and
$|3\rangle \rightarrow |4\rangle$, respectively.
$\Delta_{p}=\omega_{p}-\omega_{21}$, $\Delta_{s}=\omega_{s}-\omega_{32}$
and $\Delta_{w}=\omega_{w}-\omega_{34}$ are the detunings between the
field frequencies, $\omega_p$, $\omega_s$ and $\omega_w$, and the atomic
resonance frequencies, $\omega_{21}$, $\omega_{32}$ and $\omega_{34}$,
respectively. The Liouvillian matrix $L\left[\rho(t)\right]$ describes
relaxation by spontaneous decay, and is defined in Eq. (\ref{eq3}).
Because the ground state $|1\rangle$ and the metastable state $|4\rangle$
normally have very long coherent times (typically much larger than seconds),
here we assume there is no decay from these states.  Decay
rates $\gamma_{32}$ and
$\gamma_{34}$ give the decay from state $|3\rangle$ to states
$|2\rangle$ and
$|4\rangle$ respectively, and typically they have values of order
$10^7 s^{-1}$ or larger.  $\gamma_{21}$ is the rate from
$|2\rangle$ to
$|1\rangle$ (the intercombination transition $^{3}P_{1}
\rightarrow {^{1}S_{0}}$), and the value ranges from about
$10^6 s^{-1}$ in Yb to about $10^3 s^{-1}$ in $Ca$.  In the following
numerical calculations, we choose
$\gamma_{32}=\gamma_{34}=10\gamma_{21}=\gamma$, where $\gamma$ denotes
the atomic characteristic decay rate, and the relative value of
$\gamma_{21}$ is appropriate for Yb.  With Sr and Ca, the basic
behavior will be similar, but some numerical details will be markedly
different because $\gamma_{21}$ is much smaller.
\begin{widetext}
\begin{equation}
  L\left[\rho(t)\right]=\left[\begin{array}{cccc}
      \gamma_{21}\rho_{22} & -\gamma_{21}\rho_{12}/2 &
(\gamma_{32}+\gamma_{34})\rho_{13}/2 & 0 \\
     -\gamma_{21}\rho_{21}/2 &
-\gamma_{21}\rho_{22}+\gamma_{32}\rho_{33} &
-(\gamma_{21}+\gamma_{32}+\gamma_{34})\rho_{23}/2 &
-\gamma_{21}\rho_{24}/2 \\
     -(\gamma_{32}+\gamma_{34})\rho_{31}/2 &
-(\gamma_{21}+\gamma_{32}+\gamma_{34})\rho_{32}/2 &
-(\gamma_{32}+\gamma_{34})\rho_{33} &
-(\gamma_{32}+\gamma_{34})\rho_{34}/2  \\
      0 & -\gamma_{21}\rho_{42}/2 &
-(\gamma_{32}+\gamma_{34})\rho_{43}/2 & \gamma_{34}\rho_{33}
      \end{array}\right].
\label{eq3}
\end{equation}
\end{widetext}

\begin{figure}[h]
\includegraphics[width=5cm, height=6cm, angle=270]{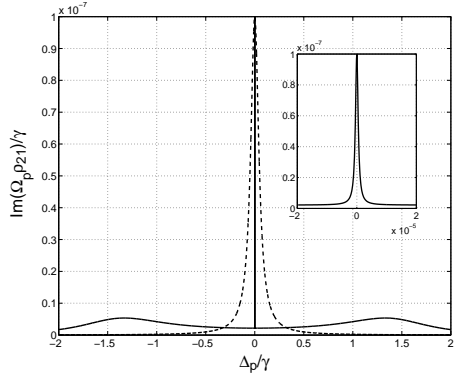}
\caption{ Absorption rate per atom of the probe light field under the
condition of electro-magnetically induced transparency and absorption
(solid line) or the normal condition of no coupling light fields (dashed
line). The inset shows a zoom-in of the sharp peak.
}
\label{fig2}
\end{figure}

Considering the system in steady state ($d\rho/dt = 0$) and retaining the
probe field only to first order in $|\Omega_{p}|^{2}$, we obtain the
following expression for the absorption rate of the probe light per atom:
\begin{eqnarray}
{\rm Im}(\Omega_{p}\rho_{21}(t))
={\rm Im}\left(\frac{|\Omega_{p}|^{2}\rho_{11}(t)}
{-2\Delta_{p}- i\gamma_{21}+ M
}\right),
\label{eq4}
\end{eqnarray}
where $M=\frac{|\Omega_{s}|^{2}}{2(\Delta_{s}+\Delta_{p})
+i(\gamma_{34}+\gamma_{32})
+\frac{|\Omega_{w}|^{2}}{2(\Delta_{w}-\Delta_{s}-\Delta_{p})}}$.
When $|\Omega_{p}|\ll |\Omega_{w}|, \gamma_{21}, \gamma_{32},
\gamma_{34} <| \Omega_{s}|$, a very sharp absorption peak
appears due to the electromagnetically induced transparency and
absorption, as shown by the solid line in Fig.~\ref{fig2}. Here for
illustration, we chose $|\Omega_{p}|=0.0001\gamma$,
$|\Omega_{w}|=0.01\gamma$,
$|\Omega_{s}|=3\gamma$, and
$\Delta_{s}=\Delta_{w}=0$.  The sharp peak is much narrower than the
normal single-photon absorption peak, shown by the dashed line in
Fig.~\ref{fig2} with
$|\Omega_{w}|=|\Omega_{s}|=0$.

\begin{figure}[h]
\includegraphics[width=5cm, height=6cm, angle=270]{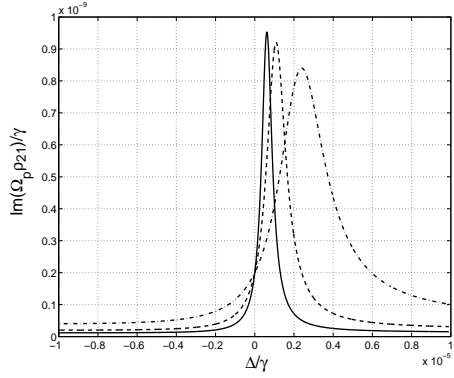}
\caption{ Narrow absorption peaks under different $|\Omega_{s}|$:
$2\gamma$ (dash-dotted line), $3\gamma$ (dashed line), or $4\gamma$
(solid line). Here $|\Omega_{p}|=0.00001\gamma$,
$|\Omega_{w}|=0.01\gamma$, $\Delta_s=0.05\gamma$, $\Delta_w=-0.05\gamma$.
}
\label{fig3}
\end{figure}

Close to the sharp absorption peak, when the three-photon detuning $\Delta\equiv
\Delta_{s}+\Delta_{p}-\Delta_{w}$ is very small (i.e. when $|\Delta| \ll
|\Omega_{w}|^{2}/|2(\Delta_{s}+\Delta_{p})+i(\gamma_{34}+\gamma_{32})|$),
Eq. (\ref{eq4}) takes a simple form that exhibits most of the important features:
\begin{eqnarray}
{\rm Im}(\Omega_{p}\rho_{21}) =
\rho_{11}|\Omega_{p}|^{2}[\gamma_{21}(1+4(\Delta -
{\bar \Delta})^{2}/W^{2})]^{-1},
\label{eq5}
\end{eqnarray}
where $W =\gamma_{21} |\Omega_{w}|^{2}/|\Omega_{s}|^{2}$ is the full
width of the three-photon resonance and ${\bar \Delta} = -\Delta_{p}
|\Omega_{w}|^{2}/|\Omega_{s}|^{2}$ is the shift in the resonance peak
from $\Delta = 0$ due to individual photon mistunings.  Both this width and
shift become arbitrarily small as
$|\Omega_{s}|$ increases, as illustrated in Fig.~\ref{fig3}. Thus the
line width of the resonance can be very narrow and give a very high
Q-value. Also, the height of the peak in Eq. (\ref{eq5}) clearly equals
the full single-photon absorption rate
$\rho_{11} |\Omega_{p}|^{2} /\gamma_{21}$, as was shown in
Fig.~\ref{fig2}.

Under the further assumption that $|\Omega_{s}/\Omega_{w}|^{2}\gg 1$,
the position of the sharp absorption peak in Eq. (\ref{eq5}) may be
written in terms of the probe laser frequency as
\begin{eqnarray}
\Delta_{p}^{Peak}=\Delta_{w}-\Delta_{s} -
W(\Delta_{w}-\Delta_{s})/\gamma_{21}.
\label{eq6}
\end{eqnarray}
This shows that the shift in the peak position from the atomic
intrinsic three-photon resonance frequency is less than the linewidth
$W$ provided the detuning of the individual fields is controlled well
enough that
$|\Delta_{s}-\Delta_{w}|< \gamma_{21}$.  In this case, if the probe
laser frequency is locked to the narrow peak, although the individual frequencies of the strong and
weak coupling laser fields might still fluctuate, the algebraic sum of the three laser frequencies is locked
very close to the three-photon resonance, i.e.,
$\omega_{p}+\omega_{s}-\omega_{w}\approx
\omega_{21}+\omega_{32}-\omega_{34}$.  External magnetic fields or
optical trapping fields can shift the intermediate
states relative to the $|1\rangle$ and $|4\rangle$ states, and
thereby shift the three-photon resonance peak a small amount according to
Eq. (\ref{eq6}).  A more detailed analysis of such effects, as well as
the effect of polarization of the individual optical fields will appear
elsewhere \cite{ToBePublished}.

\begin{figure}[h]
\includegraphics[width=5cm, height=6cm, angle=270]{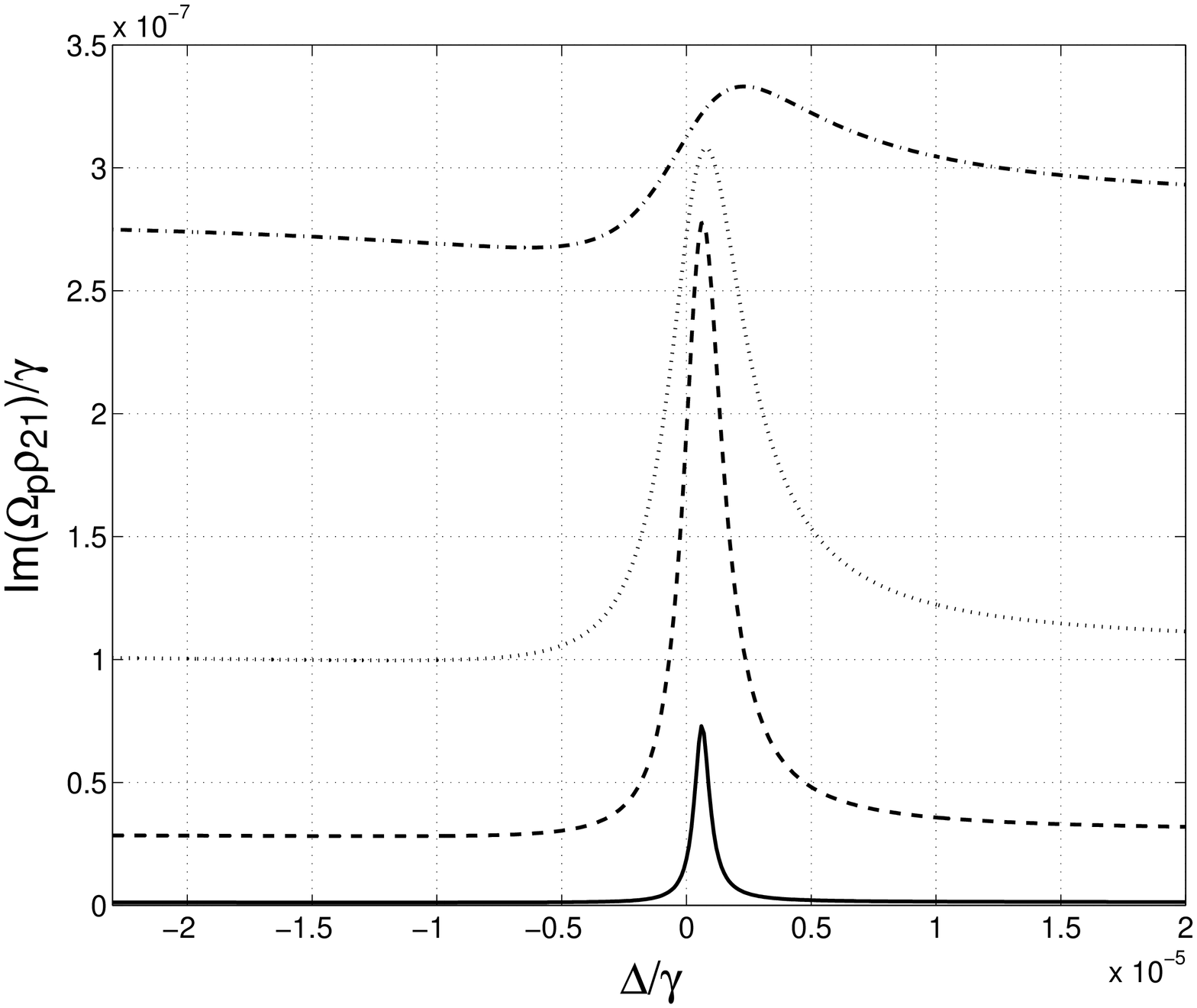}
\caption{ Narrow absorption peaks under different probe light
intensities: $|\Omega_{p}|^2=4\times 10^{-6}\gamma^2$ (dash-dotted
line), $1\times 10^{-6}\gamma^2$ (dotted line), $2.5\times
10^{-7}\gamma^2$ (dashed line), $1\times 10^{-8}\gamma^2$ (solid line).}
\label{fig4}
\end{figure}

Because the signal-noise ratio of an error signal determines the line
width of a laser locked to a frequency discriminator
\cite{LaserLineWidth}, probe light intensity is normally increased until
the absorption rate shortens the coherence time of the transition and
broadens the line.  These same considerations apply here.  As
$|\Omega_{p}|^{2}$ increases, the probed sharp absorption peak will be
broadened and, when $(\Delta_{w}-\Delta{s})$ is finite, pushed away from
the intrinsic three-photon resonance frequency, as shown in
Fig.~\ref{fig4}. Here Eq. (\ref{eq1}) is solved numerically, thereby
reflecting the nonlinear effect of
$|\Omega_{p}|^{2}$. Thus in practice, we have to make a proper trade-off between the distortion of the sharp absorption
peak and the signal magnitude. It is evident that a system composed of a large number of atoms, such as laser-cooled neutral atoms, is more ideal to realize this proposal, because the absorption rate will be increased and the low light intensity limitation can be balanced.  In addition, to reduce technical noise, some alternative ways to detect the sharp resonance feature can be considered in experiments. For example, detecting the fluorescence instead of the absorption can be considered because fluorescence peaks due to decays from states $|2\rangle$ and $|3\rangle$ have good correspondences with the absorption peak.

The Doppler effect is always a major cause of shifts and broadening
of sharp optical resonances even for cold atoms. Neutral atom
frequency standards typically use the method of Ramsey interference to
eliminate the first order Doppler effect, while trapped ion standards
and the optical lattice proposal make use of Lamb-Dicke confinement.  A
third technique becomes available for three-photon transitions:
Doppler-free alignment of the three laser beams \cite{ThreePhotonDopplerFree}.  If the
light wave vectors satisfy the phase matching relation (\ref{eq7}), the
Doppler frequency shift is zero regardless of the atomic velocity. Atoms
with arbitrary velocities can therefore contribute to the probe signal
effectively and hence enhance the signal-noise ratio. It is therefore
equivalent to the Ramsey method and superior to conventional
Doppler-free saturated absorption spectroscopy, in which only atoms with
zero velocity contribute to the signal. Also, the condition on the
wave vectors (\ref{eq8}) is relaxed in comparison with two-photon
electro-magnetically induced transparency (EIT) \cite{TwoPhotonEIT}, and
readily satisfied in Yb, Sr and Ca, as shown in Table \ref{tab:table1}.
\begin{eqnarray}
{\bf k}_{p}+{\bf k}_{s}-{\bf k}_{w}=0
\label{eq7}
\end{eqnarray}
\begin{eqnarray}
|k_{p}-k_{s}|\le k_{w}\le k_{p}+k_{s}
\label{eq8}
\end{eqnarray}
In practice, alignment of the three beams will not be perfect; however,
partial cancellation of the Doppler effect could still be very useful
because it would correspondingly increase the size of the Lamb-Dicke
region, thereby eliminating the first order Doppler effect when atoms
are confined in a region larger than the optical wavelength. An
additional advantage of this Doppler-free alignment is that there is no
net momentum transfer in the three-photon transition process from light
fields to the atom; thus there is no recoil energy shift of the
resonance.

A practical issue with the three-photon technique is how to combine the
three laser frequencies to make an optical frequency standard.  To begin
with, it is necessary for the sum of the laser frequencies,
$\omega_p+\omega_s-\omega_w$, to have a very small jitter, since this is
the effective frequency of the sharp EITA peak that constitutes the
clock reference. The complexity of independently stabilizing each laser to
its own optical cavity can be avoided by using non-linear techniques to
directly generate a beam at the clock frequency
$\omega_{clock}=\omega_p+\omega_s-\omega_w$; this
beam can be frequency-locked to a stable optical cavity by applying the
correction to only
\emph{one} laser.  The corrections applied to the single stabilized laser
will correct for both its own frequency fluctuations as well as those of
the other two lasers according to the algebraic relationship given
above. The mixing can be done either in two steps using two separate
doubly-resonant build-up cavities \cite{ShensBook} or in a single step
(as a 4-wave mixing process) in a single triply-resonant cavity
\cite{ShensBook}.  A more indirect method for
stabilizing the lasers would be to compare the three lasers to the
nearest components of a comb\cite{FSComb} generated by a femtosecond
laser whose repetition rate is stabilized to a radiofrequency source. If
$\Delta\omega_p, \Delta\omega_s$ and $\Delta\omega_w$ are the three beat
notes for the lasers, the quantity
$\Delta\omega_{clock}=\Delta\omega_p\pm\Delta\omega_s\pm\Delta\omega_w$
could be generated by radiofrequency mixing and applied to
one of the lasers using an optical modulator.  The comb spacing would be
stabilized using a fourth, ``flywheel'' laser; this would lock
the frequency of the radiofrequency source.

In conclusion, we have set forth the scheme for an optical frequency
standard based on the remarkably sharp resonance line
that appears in three-photon transitions between the $^{1}S_{0}$ and
$^{3}P_{0}$ states of alkaline earth and Yb atoms. The scheme has an
advantage of permitting the use of the even isotopes, in which the clock
transition is narrower than in the odd isotopes and should be shifted by
external magnetic fields or polarized trapping light. Based on the
electromagnetically induced transparency and absorption, the width of the
clock resonance in the scheme can be continuously adjusted from the
$MHz$ level to sub-$mHz$ without loss of signal amplitude by varying the
intensities of the three optical beams. We believe that this unique
feature will be very useful in locking lasers to the transition.
Furthermore, Doppler and recoil effects can be eliminated without the
point confinement of a lattice trap by a proper alignment of the three
beams.

Tao Hong would like to thank Yu Zhu Wang for helpful discussion. This
work was  supported by the National Science Foundation, Grant No. PHY
0099535.

\newpage
\begin{table}
\caption{\label{tab:table1} Atom and corresponding optical wavelength
candidates for forming the scheme in Fig.~\ref{fig1}. }
\begin{ruledtabular}
\begin{tabular}{l c c c c}
Atom & Probe & Strong & Weak & Clock  \\
&($^{1}S_{0}$-$^{3}P_{1}$) & ($^{3}P_{1}$-$^{3}S_{1}$)
&($^{3}S_{1}$-$^{3}P_{0}$) & \\ \hline
Yb & 556 nm & 649 nm & 680 nm & 578 nm \\ \hline
Sr & 689 nm & 688 nm & 679 nm & 698 nm \\ \hline
Ca & 657 nm & 612 nm & 610 nm & 659 nm \\
\end{tabular}
\end{ruledtabular}
\end{table}

\begin{thebibliography}{99}
\bibitem{TwoPhotonEIT} In a two-photon EIT system, the Doppler effect cannot be completely eliminated unless the two light fields are degenerate in frequency.
\end{thebibliography}
\end{document}